# Changing the Phosphorus Allotrope from a Square Columnar Structure to a Planar Zigzag Nanoribbon by Increasing the Diameter of Carbon Nanotube Nanoreactors

Jinying Zhang,*[†] Chengcheng Fu,[†] Shixin Song,[‡] Hongchu Du,[§] Dan Zhao,[†] Hongyang Huang,[†] Lihui Zhang,[†] Jie Guan,*[‡] Yifan Zhang,[⊥] Xinluo Zhao,[⊥] Chuansheng Ma,[∥] Chun-Lin Jia,*[§, ∥] David Tománek[#]

[†]State Key Laboratory of Electrical Insulation and Power Equipment, Center of Nanomaterials for Renewable Energy, School of Electrical Engineering, Xi'an Jiaotong University, Xi'an, Shaanxi, P. R. China, 710049

[‡]The School of Physics, Southeast University, Nanjing, Jiangsu, P. R. China, 211189

[§]Ernst Ruska Centre for Microscopy and Spectroscopy with Electrons, Forschungszentrum Jülich, 52425 Jülich, Germany

[⊥]Department of Physics, Shanghai University, Shanghai, P. R. China, 200444

∥ The School of Microelectronics and State Key Laboratory for Mechanical Behavior of Materials, Xi'an Jiaotong University, Xi'an, Shaanxi, P. R. China, 710049

#Physics and Astronomy Department, Michigan State University, East Lansing, Michigan 48824-2320, USA

ABSTRACT. Elemental phosphorus nanostructures are notorious for a large number of allotropes, which limits their usefulness as semiconductors. To limit this structural diversity, we synthesize selectively quasi-1D phosphorus nanostructures inside carbon nanotubes (CNTs) that act both as stable templates and nanoreactors. Whereas zigzag phosphorus nanoribbons form preferably in CNTs with an inner diameter exceeding 1.4 nm, a previously unknown square columnar structure of phosphorus is observed to form inside narrower nanotubes. Our findings are supported by electron microscopy and Raman spectroscopy observations as well as *ab initio* density functional theory calculations. Our computational results suggest that square columnar structures form preferably in CNTs with inner diameter around 1.0 nm, whereas black phosphorus nanoribbons form preferably inside CNTs with 4.1 nm inner diameter, with zigzag nanoribbons energetically favored over armchair nanoribbons. Our theoretical predictions agree with the experimental findings.

Layered black phosphorus and the monolayer structure, dubbed phosphorene,[1] have attracted much attention due to their tuneable semiconducting character.[2] Phosphorus has a rich phase diagram that is still evolving. White/yellow phosphorus,[3, 4] consisting of $P_4$ molecules, is a well known allotrope. Black phosphorus, consisting of planar hexagonal structure with armchair ridges, was first synthesized in 1914 by Percy W. Bridgman.[5] Fibrous phosphorus, composed of -[P2]-[P8]-[P2]-[P9] tubular strands, was first produced in 2005.[6] In this allotrope, all strands are parallel to each other, with pairs of parallel strands covalently linked through [P9]. Violet phosphorus, also called Hittorf phosphorus, consists of layers of parallel -[P2]-[P8]-[P2]-[P9] tubular strands. Even though adjacent layers are rotated in-plane by 90°, strands in adjacent layers are still covalently linked through [P9]. This structure was first synthesized in 1865[7] and its lattice structure was subsequently described by Thurn and Krebs in 1969.[8] Single crystals of violet phosphorus have recently been produced and their lattice structure has been obtained reliably by single-crystal X-ray diffraction.[9] In addition to the experimentally well established white/yellow phosphorus, black phosphorus, violet phosphorus, and fibrous phosphorus allotropes, various other phosphorus structures were also predicted.[10-13] Properties of phosphorus depend to a high degree on the allotrope structure. Unique photoelectronic properties are emerging as the size of the structure decreases. Different phosphorus allotrope nanostructures are also very interesting candidates for photoelectronic materials, but their properties vary from allotrope to allotrope. Adding to this problem, selective synthesis of specific phosphorus allotropes, especially on the nanoscale, remains a formidable challenge. Carbon nanotubes have been demonstrated to act as effective templates and nanoreactors to synthesize and stabilize particular meta-stable nanostructures.[14-24]

Here, we postulate a previously unknown square columnar phosphorus structures (SC-P) and compare their stability to that of nanoribbons. SC-P structures form inside single-walled carbon nanotubes (SWCNTs) with inner diameter around 1.0 nm, whereas zigzag nanoribbons of phosphorus (ZR-Ps) have been observed to form inside wider multi-walled carbon nanotubes (MWCNTs) with inner diameter of 4.1 nm.

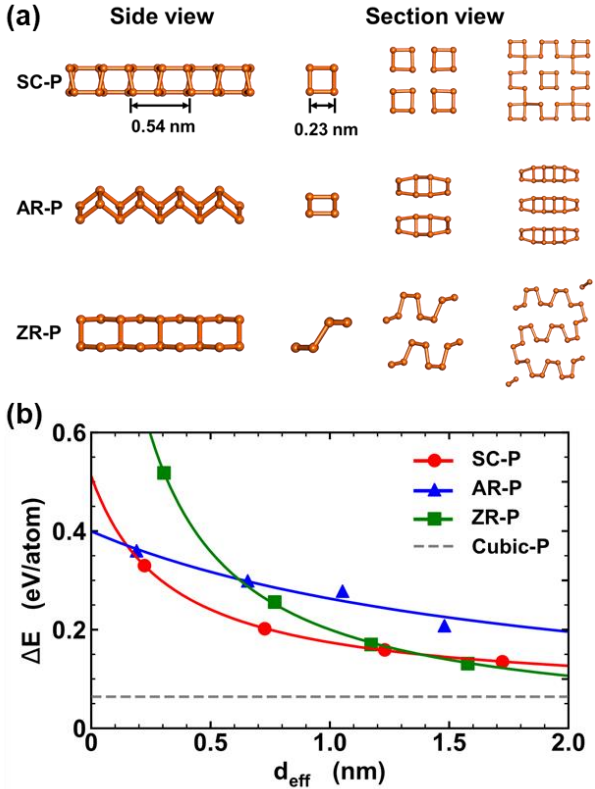

**Figure 1.** (a) Structural models of possible 1D phosphorus allotropes with different diameters. The left two columns are the side view and section view of isolated SC-P, the narrowest AR-P and ZR-P. The right two columns are the section view of 2×2, 3×3 SC-P arrays and the corresponding AR-P, ZR-P multilayers with comparable diameters. (b) The relative energy ΔE as a function of the effective diameter $d_{eff}$ for 1D SC-P, AR-P and ZR-P structures. The total energy of bulk black phosphorus is set to zero as a reference and that of cubic phosphorus is shown by the horizontal dashed line.

Finite-size allotropes of systems like black P, which prefer planar geometries, are characterized by the competition between the bending energy and the edge energy. When the energy to form exposed edges exceeds the bending energy, we expect formation of tubular structures with no edges. In the opposite case, we expect formation of planar finite-width ribbons. Formation of one or the other structure may be favored selectively by steric constraints such as the finite void inside a carbon nanotube.

Considering different structures computationally, we predict that a previously unknown 1D SC-P structure may be stabilized inside a narrow carbon nanotube. When comparing the equilibrium structures shown in the first row of Figure 1a, an isolated SC-P is essentially a narrow nanotube with a square cross section. The distorted P squares are connected to each other by two P-P bonds along the columnar axis and there are eight P atoms in one unit cell. The side length of the cross section is 0.23 nm and the lattice constant along the columnar axis is 0.54 nm. Several SC-Ps in parallel may coexist in carbon nanotubes with larger inner diameters. The energetic stability of n×n (n $\geq$ 1) SC-P arrays is compared to those of armchair nanoribbons of phosphorus (AR-P) and ZR-P monolayers and multilayers, which may also form inside carbon nanotubes. Focusing on the equilibrium structures in the last two rows of Figure 1a, the narrowest AR-P and ZR-P structures contain four atoms in the section view, similar to an isolated SC-P. Multilayers of wider AR-Ps and ZR-Ps are expected to exist inside carbon nanotubes with larger inner diameters.

The relative energy ($\Delta E$) with respect to bulk black phosphorus (BP) for structures from the narrowest isolated SC-P up to a 4×4 SC-P array, along with the energies of AR-Ps and ZR-Ps with comparable sectional sizes, have been calculated using density functional theory (DFT). The first impression of our results in Figure 1b is that $\Delta E$ decreases with increasing effective diameter $d_{eff}$ for the three types of 1D structures. In our study, we estimated $d_{eff}$ from the average of the sectional side lengths along different directions. When $d_{eff}$ is smaller than 1.4 nm, SC-P is shown to be the most stable structure. However, ZR-P becomes the most stable allotrope when $d_{eff}$ is larger than 1.4 nm. It is worth noting that if there is no spatial constraint in our theoretical study, all 1D structures shown here will transform to 3D structures. Both AR-P and ZR-P will transform to bulk BP, which is energetically the most stable allotrope for $d_{eff} \rightarrow \infty$. Under similar conditions, SC-P will transform to a cubic phase of phosphorus that is energetically only 0.06 eV/atom less stable than bulk BP, as shown by the horizontal dashed line in Figure 1b.

Energetic preference for the phosphorus structures discussed here was confirmed experimentally in carbon nanotubes with diameters smaller than 5 nm. In wider MWCNTs with inner diameters around 5-8 nm, ring/coil-shaped phosphorus allotropes have been observed[22] and shown to be preferred energetically.[25] SWCNTs used in this study were prepared using an arc discharge[26] method and MWCNTs were formed using a chemical vapor deposition (CVD) method.[27] Encapsulation of phosphorus into SWCNTs and MWCNTs was achieved by a vapor reaction method at 500ºC using amorphous phosphorus as precursor. This approach has been described in the synthesis of ring/coil-shaped phosphorus previously.[22]

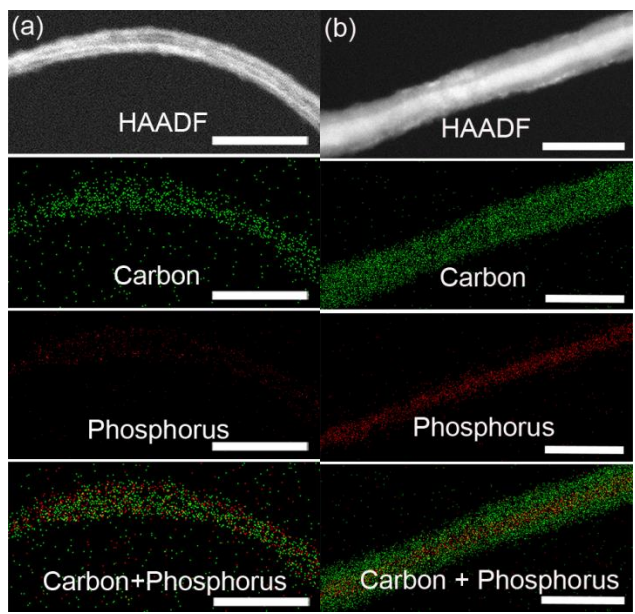

**Figure 2.** Representative HAADF images and corresponding elemental maps of C, P and C+P of phosphorus encapsulated inside (a) SWCNTs with inner diameters around 1 nm and (b) MWCNTs with inner diameters around 4 nm. (scale bar = 25 nm) (JEOL JEM-F200 (HR), accelerating voltage: 200 kV)

Encapsulation of phosphorus into SWCNTs and MWCNTs has been confirmed by scanning transmission electron microscopy (STEM) and elemental analysis. Encapsulation of atoms heavier than carbon inside carbon nanotubes has been confirmed by the brighter image contrasts in the high angle annular dark-field (HAADF) images that place such atoms inside the surrounding carbon nanotube walls, as seen in Figure 2. The encapsulated structures in both SWCNTs and MWCNTs were confirmed to be phosphorus by elemental maps of P and C. SWCNT bundles with extremely thin brighter phosphorus nanocolumns were easily observed in the HAADF of Figure 2a, while the phosphorus nanocolumns in MWCNTs are much thicker as shown in the HAADF images of Figure 2b. The HAADF images of encapsulated SWCNT bundles instead of isolated SWCNTs were used since the isolated SWCNTs containing phosphorus were easily broken during focusing. STEM and elemental analysis results indicate clearly that phosphorus has been encapsulated inside the innermost voids of carbon nanotubes with different inner diameters.

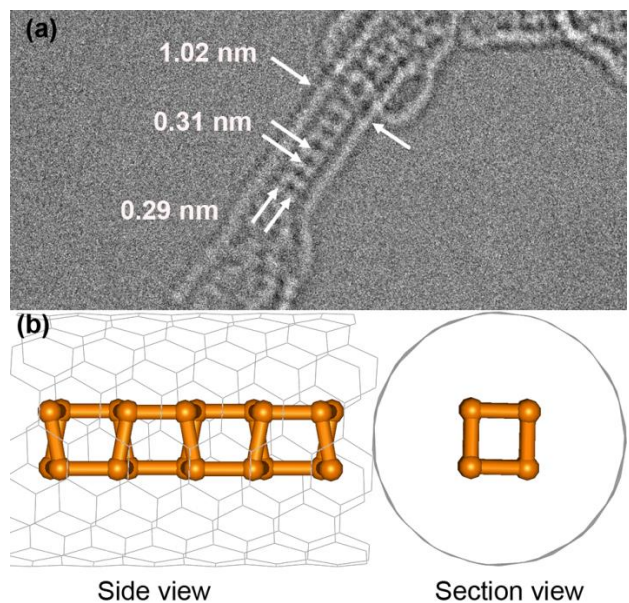

**Figure 3.** (a) NCSI image of SC-P@SWCNT (FEI Titan³ 60-300 (PICO) microscope equipped with a spherical-chromatic aberration corrector, accelerating voltage: 80 kV). (b) Structural model of SC-P@SWCNT.

SC-Ps were observed to form inside SWCNTs with inner diameters around 1.0 nm, which is consistent with our finding in Figure 1b that SC-P is the most stable structure of phosphorus with intersectional size below 1.4 nm. The structure was investigated by the negative spherical aberration imaging (NCSI) technique.[28, 29] The NCSI image of SC-P@SWCNT in Figure 3a was recorded at 80 kV. Atoms were observed to be bright against a dark background. The encapsulated structure inside the SWCNT is well-resolved and understood on the basis of our calculations. Based on the high resolution transmission electron microscope (HRTEM) image of the SC-P inside a SWCNT with diameter of 1.02 nm, presented in Figure 3a, the width of the connected square units in the SC-P is 0.29 nm and the distance between two connected square units along the SC-P axis is 0.31 nm. The in-plane distance between two connected phosphorus atom units is slightly larger than that of SC-P@SWCNT (0.27 nm), but much larger than that of AR-P@SWCNT (Figure S1a, 0.17 nm) or ZR-P@SWCNT (Figure S1b, 0.16 nm). The structure shown in HRTEM image is also consistent with the shape of SC-P@SWCNT, rather than that of AR-P@SWCNT or ZR-P@SWCNT. The HRTEM image of the SC-P inside the SWCNT is consistent well with the side view of SC-P@SWCNT shown in Figure 3b. The size of the observed SC-P in the HRTEM image is slightly larger than that of the theoretically predicted SC-P. This may be partly due to the attractive interaction between the phosphorus nanocolumn and the surrounding SWCNT, which has not been considered in the calculation and which may cause structural deformations.

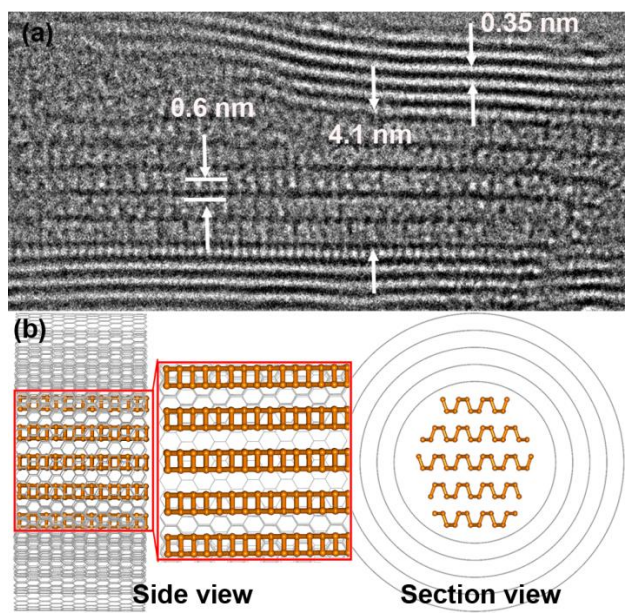

**Figure 4.** (a) NCSI image of a ZR-P@MWCNT (FEI Titan³ 60-300 (PICO) microscope equipped with a spherical-chromatic aberration corrector, accelerating voltage: 80 kV). (b) Structural model of ZR-P@MWCNTs.

Increasing the inner diameter of the carbon nanotubes, we observed formation of ZR-Ps instead of SC-Ps. We were also able to observe by HRTEM the structure of the ZR-Ps encapsulated inside MWCNTs with inner diameter of 4.1 nm. In the side view, the observed encapsulated phosphorus structures are consistent with calculated ZR-Ps inside carbon nanotubes. As seen in the NCSI image in Figure 4a, the measured interlayer distance of 0.6 nm agrees well with the theoretically predicted value for a stack of ZR-Ps inside a MWCNT with inner diameter of 4.1 nm presented in Figure 4b. We note at this point that the ZR-P structure inside MWCNTs with inner diameter of 4.1 nm is distinguishable from the ring/coil-shaped structures.[22] Whereas the ZR-P structure is aligned along the axis of the carbon nanotube, the structure of the ring/coil-shaped phosphorus coils around the inner diameter of the carbon nanotube and thus appears perpendicular to the nanotube axis.[22, 25]

The demonstrated successful synthesis of SC-Ps in SWCNTs with inner diameter of 1.0 nm and ZR-Ps in MWCNTs with inner diameter of 4.1 nm is in very good agreement with the theoretical prediction that SC-Ps are the most stable structures with intersectional sizes below 1.4 nm and ZR-Ps are the most stable structures with intersectional sizes beyond 1.4 nm. Both SC-Ps and ZR-Ps were observed to be stable while encapsulated, since the lone pair electrons of phosphorus are passivated by the intact walls of carbon nanotubes.

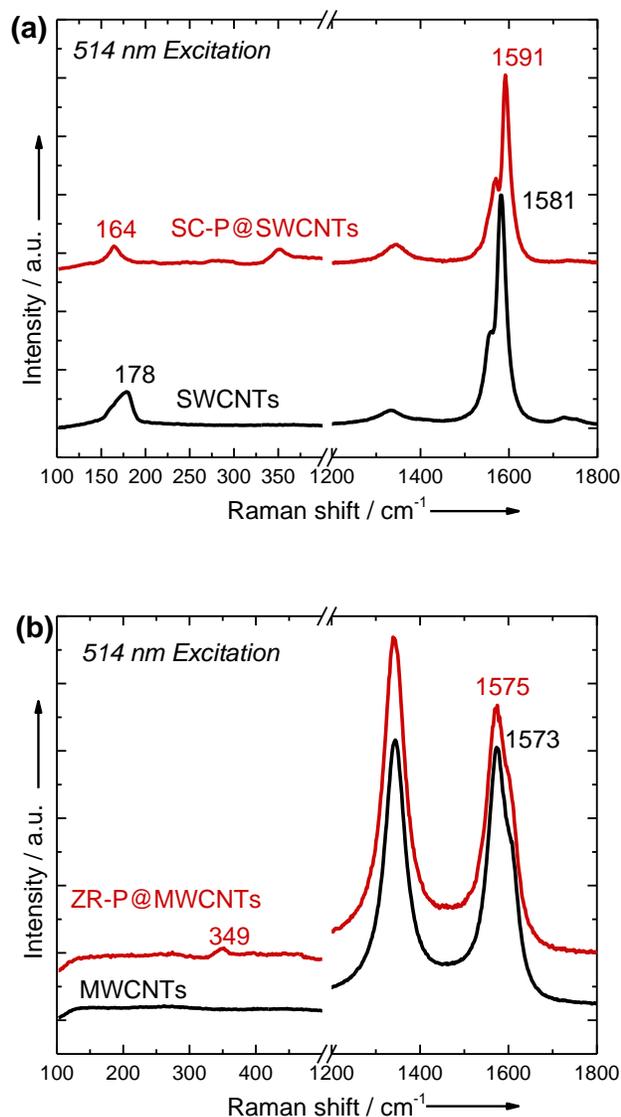

**Figure 5.** Raman spectra of pristine and phosphorus-filled (a) SWCNTs and (b) MWCNTs. Spectra of pristine nanotubes are shown in black and those of P-filled nanotubes in red.

We also used Raman scattering to characterize SC-P@SWCNTs and ZR-P@MWCNTs and compare the spectra to those of empty SWCNTs and MWCNTs. The Raman results are presented in Figure 5. Only a fraction of the as-decomposed red phosphorus clusters entered into the void inside carbon nanotubes to form SC-Ps and ZR-Ps. The remaining fraction of the sublimed red phosphorus condensed outside the carbon nanotubes, forming an amorphous phase and crystalline red phosphorus during the vapor phase reaction at 500ºC. It turned out to be a significant challenge to completely remove the red phosphorus condensed outside carbon nanotubes even during an acetone wash. The Raman features between 340–600 cm$^{-1}$ originate from deposited red phosphorus on the external surfaces of carbon nanotubes, as shown by the broad bands in Figure 5, and will not be discussed any more in this study. It is hard to detect P-P vibrations of the encapsulated phosphorus structures inside carbon nanotubes due to the laser beam absorption by the carbon nanotubes walls, which is consistent with previously reported results for ring/coil-shaped phosphorus.[22] However, a blue shift has been detected for the G-band of both SWCNTs and MWCNTs after encapsulation. We interpret the cause being hole doping of the nanotubes by encapsulated phosphorus, in agreement with reported data.[30] This blue shift in the SWCNT G-band is significantly larger than the corresponding shift in the MWCNTs, where the doping level decreases with increasing inner tube diameter. The RBM band of the SWCNTs was also observed be red shifted by about 6 cm$^{-1}$ after encapsulation of phosphorus due to the coupling between the surrounding carbon nanotube walls and the enclosed phosphorus structures. No Raman feature corresponding to the stretching modes of P-C bond (650-770 cm$^{-1}$)[31] was observed for encapsulated SC-P@SWCNTs or ZR-P@MWCNTs, as seen in Figure S2 in the SI.

As a previously unexplored allotrope of phosphorus, the dynamic stability and electronic structure for the 1D free-standing SC-P has been investigated by our DFT calculations. According to the phonon spectrum results shown in Figure S3 in the SI, a straight isolated SC-P structure is unstable, as indicated by one imaginary frequency. We found a dynamically stable structure of SC-P, in which the two connected squares in one unit cell are slightly rotated around the columnar axis relative to each other, as seen in Figure S3b. This slight twist lowers the total energy of SC-P by 0.03 eV/atom and yields a phonon spectrum with no imaginary frequencies. The lattice constant for the twisted SC-P is almost the same as the original untwisted structure. Our DFT-PBE electronic structure results for the stable twisted SC-P, presented in Figure S4 in the SI, indicate an indirect-gap of 1.88 eV, which is nearly twice as large as that of phosphorene (1.0 eV)[1]. The semiconducting character of SC-P indicates its potential use in 1D electronics and optoelectronics.

In summary, we have predicted a previously unknown phosphorus square columnar (SC-P) structure with the intersectional size around 1.4 nm that should be stable inside carbon nanotubes with inner diameter smaller than 2.1 nm. The SC-P structure was produced and observed inside SWCNTs with inner diameter around 1 nm, which acted as a template and nanoreactor. NCSI images of the square column indicate a width of 0.29 nm and inter-square distance of 0.31 nm along the columnar axis. These values agree well with the theoretical prediction. The fact that the observed structural values are slightly larger than the computed ones can be attributed to the attractive interaction between SC-P and the wall of the SWCNT. Planar zigzag black phosphorus ribbons are predicted theoretically to be more stable than SC-P and armchair black phosphorus ribbons. Based on their stability, they should dominate among structures with intersectional sizes exceeding 1.4 nm and may be contained in carbon nanotubes with inner diameter around 2.1 nm. Also this prediction has been confirmed by the observation of zigzag black phosphorus ribbons inside MWCNTs with inner diameter of 4.1 nm. The phosphorus square columns and zigzag black phosphorus ribbons are distinguishable from the ring/coil-shaped phosphorus structures previously reported inside MWCNTs with inner diameters around 5-8 nm. Thus, tuning the diameter of the carbon nanotube nanoreactor leads to a selective synthesis of the SC-P allotrope in very narrow nanotubes, the ZR-P allotrope in wider nanotubes, and ring/coil structures in nanotubes with diameters between 5-8 nm.

**Experimental Section**

SWCNTs and MWCNTs were produced by the arc discharge method and chemical vapor deposition (CVD), respectively. The SWCNTs were heated to 420°C and the MWCNTs to 500°C for 30 min under air atmosphere to open the caps. The open-ended carbon nanotubes were degassed for one day with interval flame heating and then sealed in the presence of extra red phosphorus (Aladdin, 99.999% metals basis) under a vacuum of $10^{-5}$ Pa in an H-shaped Pyrex tube. The H-shaped Pyrex tube was then heated to 500°C for 48 h with a rate of 1°C/min and then cooled down in the oven. Raman spectroscopy was taken in a back-scattering geometry using a single monochromator with a microscope (Reinishaw inVia) equipped with a CCD array detector (1024 × 256 pixels, cooled to -70ºC) and an edge filter. The Raman spectra were recorded with an excitation laser at 514 nm. The spectral resolution and reproducibility was determined to be better than 0.1 cm$^{-1}$. High-resolution transmission electron microscopy investigations were carried out at an acceleration voltage of 80 kV on an FEI Titan[3] 60-300 (PICO) microscope equipped with a high-brightness field emission gun, a monochromator, and a $C_S$-$C_C$ (spherical-chromatic aberration) achro-aplanat image corrector for the objective lens, providing an attainable resolution of 80 pm[32]. HAADF and elemental mapping of C and P was obtained using the JEOL JEM-F200 (HR) transmission electron microscope.

**Theoretical Methods**

*Ab initio* density functional theory (DFT) was utilized as implemented in the Vienna ab initio Simulation Package (VASP)[33] to obtain the optimized structure and total energies for the phosphorus structures of interest. The Perdew-Burke-Ernzerhof[34] (PBE) exchange correlation functional and projector-augmented-wave[35,36] (PAW) pseudopotentials were used. Van der Waals interactions have been considered by introducing the DFT-D2[37] correction. An energy cut-off of 500 eV for the plane wave basis was used and a criterion of $10^{-5}$ eV was set for the maximum total energy difference between subsequent self-consistency iterations at the point of self-consistency. All geometries have been optimized using the conjugate-gradient method,[38] until none of the residual Hellmann-Feynman forces exceeded the threshold of $10^{-2}$ eV/ Å. Periodic boundary conditions were used and all 1D structures were separated by a vacuum region in excess of 20 Å. A 1×1×8 k-point grid[39] was applied in the Brillouin zone of the 1D SC-P structures or its equivalent in that of unit cells with different sizes.

ASSOCIATED CONTENT

**Supporting Information**.

An associated pdf file containing details of the experiment and calculation is provided free of charge.

AUTHOR INFORMATION

**Corresponding Authors**

*Email: jinying.zhang@mail.xjtu.edu.cn

*Email: guanjie@seu.edu.cn

*Email: c.jia@fz-juelich.de or c.jia@mail.xjtu.edu.cn

**Funding Sources**

This work has been supported by the National Natural Science Foundation of China (21771143 and 61704110), the U.S. National Science Foundation and the AFOSR EFRI 2-DARE program (EFMA-1433459).

**Notes**

The authors declare no competing financial interest.

ACKNOWLEDGMENT

J.Z. is supported by the CyrusTang Foundation through the Tang Scholar Program. J.G. is supported by Fundamental Research Funds for the Central Universities, the Shuangchuang Doctor Program of Jiangsu Province, and the Southeast University "Zhongying Young Scholars" Project. H.D. acknowledges the support from the Deutsche Forschungsgemeinschaft (DFG) under grant of the SFB 917 "Nanoswitches". D.T. acknowledges financial support by the NSF/AFOSR EFRI 2-DARE grant number EFMA-1433459.

ABBREVIATIONS

SC-P, square columnar phosphorus structures; ZR-P, zigzag nanoribbon of phosphorus; AR-P, armchair nanoribbon of phosphorus; SWCNTs, single-walled carbon nanotubes; MWCNTs, multi-walled carbon nanotubes; BP, black phosphorus; DFT, density functional theory; CVD, chemical vapor deposition; STEM, scanning transmission electron microscopy; HAADF, high angle annular dark-field; NCSI, negative $C_S$ (spherical aberration) imaging; HRTEM, high resolution transmission electron microscope.

SYNOPSIS (Word Style "SN_Synopsis_TOC").

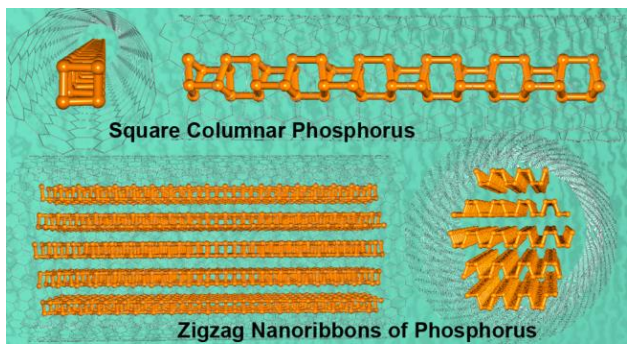

# Changing the Phosphorus Allotrope from a Square Columnar Structure to a Planar Zigzag Nanoribbon by Increasing the Diameter of Carbon Nanotube Nanoreactors

*Jinying Zhang,*[*,†] *Chengcheng Fu,*[†] *Shixin Song,*[‡] *Hongchu Du,*[§] *Dan Zhao,*[†] *Hongyang Huang,*[†] *Lihui Zhang,*[†] *Jie Guan,*[*,‡] *Yifan Zhang,*[⊥] *Xinluo Zhao,*[⊥] *Chuansheng Ma,*[//] *Chun-Lin Jia,*[*,§,//] *David Tománek*[#]

[†]State Key Laboratory of Electrical Insulation and Power Equipment, Center of Nanomaterials for Renewable Energy, School of Electrical Engineering, Xi'an Jiaotong University, Xi'an, Shaanxi, P. R. China, 710049

[‡]The School of Physics, Southeast University, Nanjing, Jiangsu, P. R. China, 211189

[§]Ernst Ruska Centre for Microscopy and Spectroscopy with Electrons, Forschungszentrum Jülich, 52425 Jülich, Germany

[⊥]Department of Physics, Shanghai University, Shanghai, P. R. China, 200444

[//] The School of Microelectronics and State Key Laboratory for Mechanical Behaviour of Materials, Xi'an Jiaotong University, Xi'an, Shaanxi, P. R. China, 710049

[#]Physics and Astronomy Department, Michigan State University, East Lansing, Michigan 48824-2320, USA

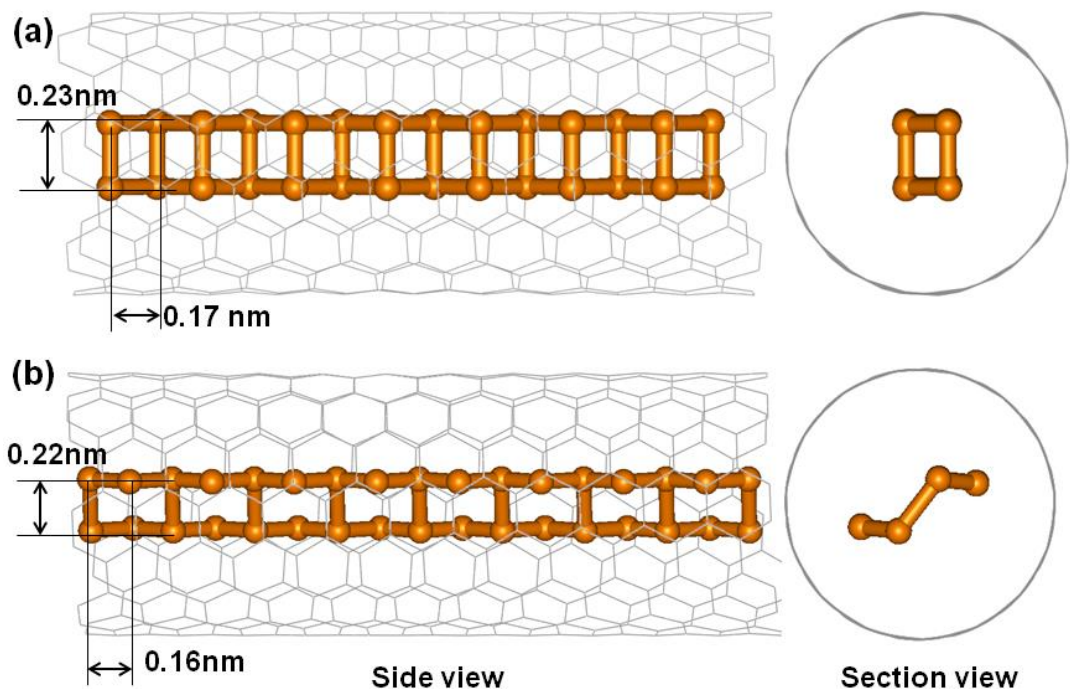

**Figure S1.** Structural models of (a) AR-P@SWCNT and (b) AR-P@SWCNT.

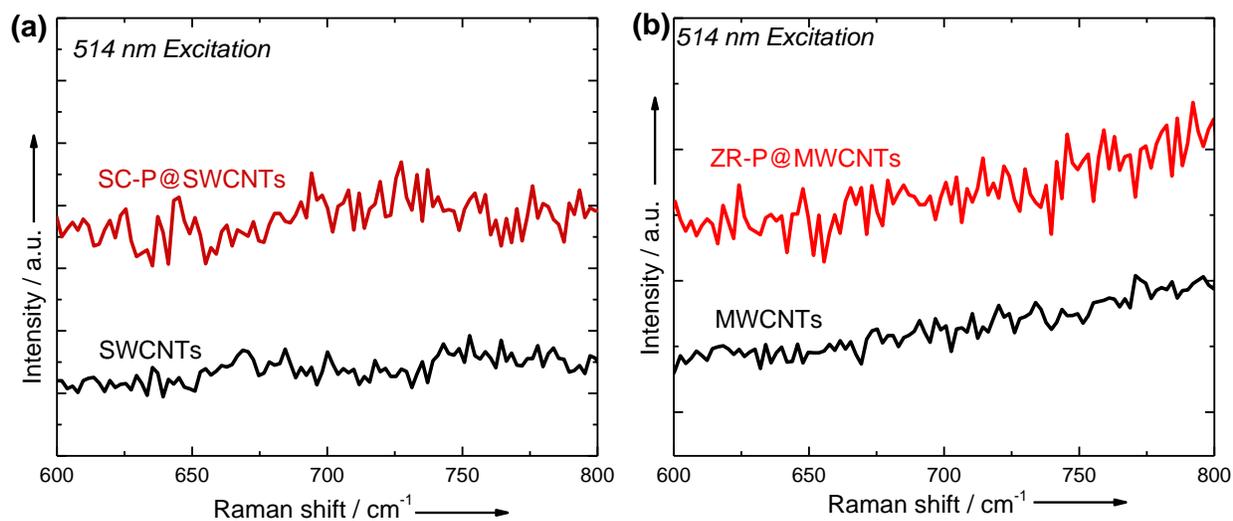

**Figure S2.** Raman spectra of pristine and phosphorus-filled (a) SWCNTs and (b) MWCNTs. Spectra of pristine nanotubes are shown in black and those of P-filled nanotubes in red.

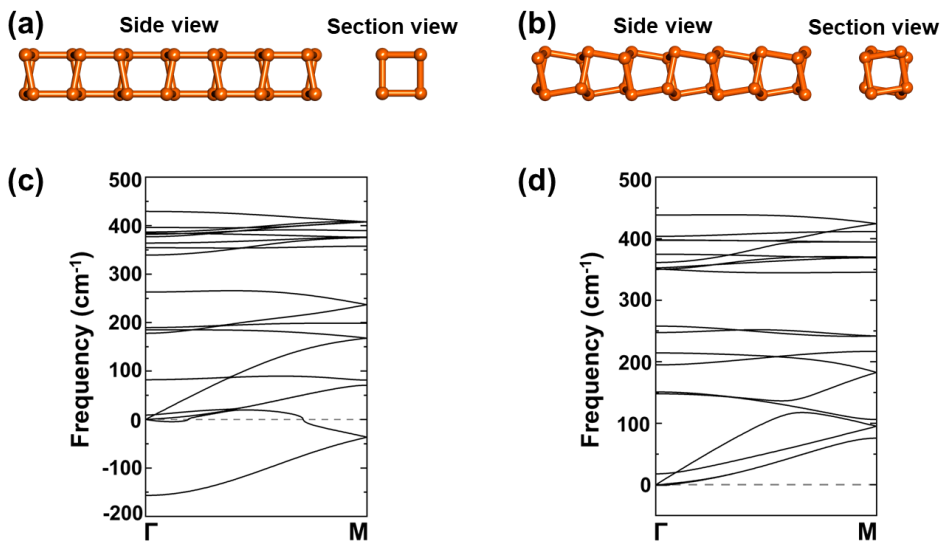

**Figure S3.** (a, b) structural model and (c, d) vibrational phonon spectra of unstable straight (left) and stable twisted (right) SC-P structures.

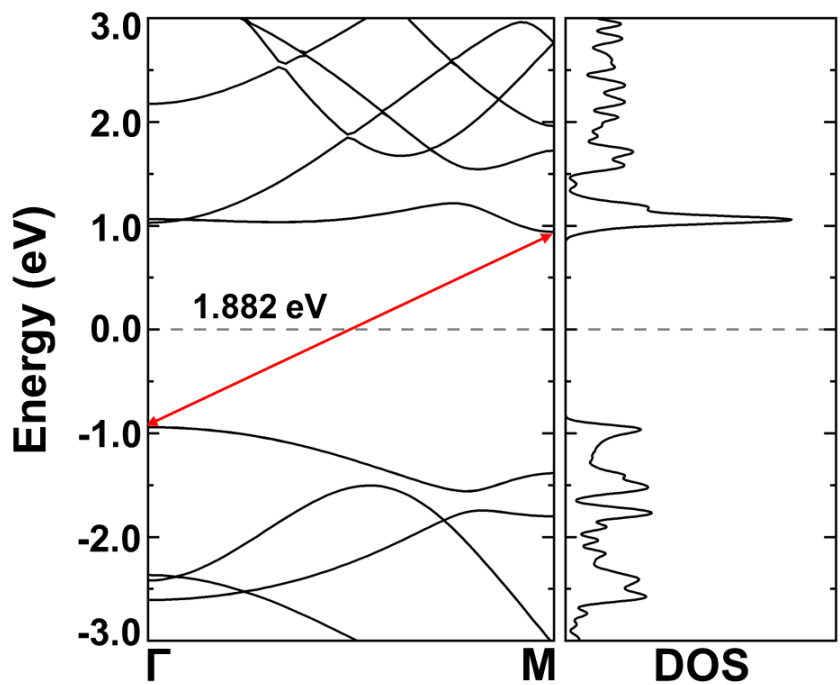

**Figure S4.** Electronic band structure and density of states (DOS) for the twisted SC-P allotrope based on DFT-PBE calculations. An indirect band gap is indicated by the red arrow.